# DESIGN AND IMPLEMENTATION VOIP SERVICE ON OPEN IMS AND ASTERISK SERVERS INTERCONNECTED THROUGH ENUM SERVER

Rendy Munadi [1], Effan Najwaini [2], Asep Mulyana [3], R.Rumani.M [4]

Telkom Institute of Technology , Bandung 40257 West Java-INDONESIA

Email : rnd@ittelkom.ac.id[1] effan_86@yahoo.com[2],
asm@ittelkom.ac.id[3], rrm@ittelkom.ac.id[4]

## ABSTRACT

*Asterisk and Open IMS use SIP signal protocol to enable both of them can be connected. To facilitate both relationships, Enum server- that is able to translate the numbering address such as PSTN (E.164) to URI address (Uniform Resource Identifier)- can be used. In this research, we interconnect Open IMS and Asterisk server Enum server. We then analyze the server performance and PDD (Post Dial Delay) values resulted by the system.*

*As the result of the experiment, we found that, for a call from Open IMS user to analog Asterisk telephone (FXS) with a arrival call each servers is 30 call/sec, the maximum PDD value is 493.656 ms. Open IMS is able to serve maximum 30 call/s with computer processor 1.55 GHz, while the Asterisk with computer processor 3.0 GHz, may serve up to 55 call/sec. Enum on server with 1.15 GHz computer processor have the capability of serving maximum of 8156 queries/sec.*

## 1. INTRODUCTION

Interconnection and convergence among PSTN, PLMN and data network are expected to develop very powerful system; this system will provide PSTN services, mobility, PLMN features and internet-based application. This convergence will support multimedia service with adequate bandwidth and high mobility. The IMS technology is resulted from combination of multimedia, mobile and IP concepts in order to complete the NGN technology [3,4].

.IMS acts as a standard platform for multimedia service through IP/SIP protocol with which it is possible for the operator to use one's platform for many multimedia service. IMS is a part of arhitecture standard of *Next Generation Network* (NGN). Services for *fixed, mobile and wireless networks* can be operated through IMS platform with IP based service and supported by SIP protocol [1,5].

.Originally, IMS is developed for telephone connection in mobile network, but together with TISPAN release-7, it is also has the possibility for fixed network, resulting term *Fixed-Mobile Convergence* (FMC) which is a key industries trend in 2005 [5].

Software based on NGN architecture has been developed today. OpenIMS is constitute of the software based on IMS architecture; it has the capability for various multimedia features.On the other hand, Asterisk is a software based on softswitch architecture which is capable to connect packet network and circuit network. With those both software, we can build a simple NGN technology with lower cost investment.

To demonstrate the above cases, we design and implement a VoIP service, using the above software, with the aid of Enum server which the function is to mapping the number among servers.

## 2. BASIC CONCEPT
### 2.1. Open IMS

OpenIMS (Open Source IMS) is a software made by FOKUS (Germany Institute) on December 2006. FOKUS implemented in an integrated way for some IMS components, such as CSCFs, HSS, Application *Server*s and others, which can be seen in Figure.1. This software has different platform services, such as "Open Service Access (OSA)/Parlay, JAIN Service Logic Execution Environment (SLEE), Web services/Parlay X, SIP Servlets, Call Processing Language (CPL)" [16].
.

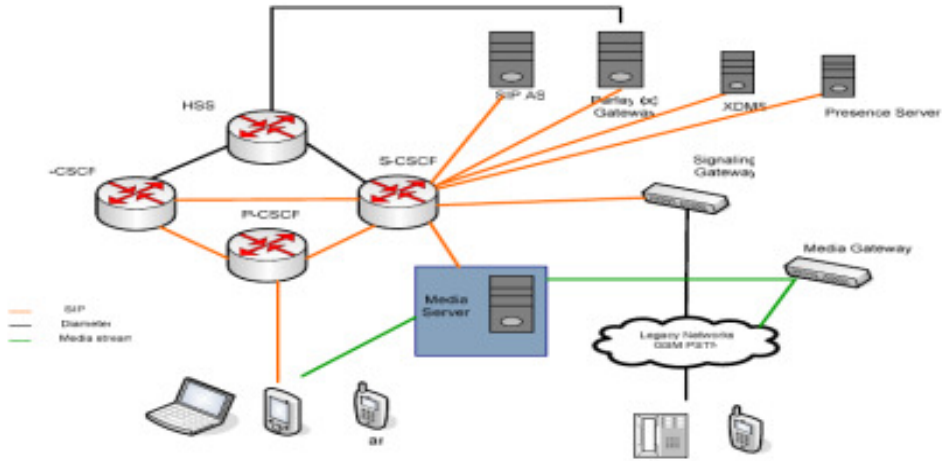

Figure 1  OpenIMS Network Overview

## 2.2. Asterisk

Asterisk is an open source software which can be used to develop communication services. Asterisk it self provides simplicity for the users to improve telephone services by themselves with flexible customization by users. Open source means that the developer can change source codes so the applications can be added easily by the developer.

Asterisk can be considered as a complete PBX or Software complete PBX and provide all PBX features. The advantage of Asterisk: can run under some OS such Linux, Windows, BSD and OS X. Another thing is, it can be connected with almost all of POTS standard with minimum/ cheap additional hardware as a gateway. Some features are provided such as Voicemail, Conference Call, Interactive Voice Response, Call Queuing, Three Way Calling, Caller ID Services, Analog Display Service Interface, SIP VoIP Protocol,  H323 (as a client & gateway), IAX, MGCP (provide call manager function only), SCCP/Skinny, and others [2].

The title is to be written in 20 pt. Garamond font, centred and using the bold and "Small Caps" formats.  There should be 24 pt. (paragraph) spacing after the last line.

## 2.3. Enum Overview

Electronic Number Mapping (ENUM) is a number mapping mechanism. Numbers mapping means Electronic Equipments identity numbers (ITU-T E.164 recommendation) to DNS Uniform *Resource* Identifier (*URI*) system [10]. The DNS URI has been globally used in Internet. *ENUM* is a standard recommended by IETF (RFC 3761). It allow user to use a phone numbers to access DNS. Then, user can access *record Universal Resource Identifier* inside *NAPTR Resource Record* which is belong to its numbers, which is can be seen in Figure.2.

Electronic Number Mapping is a mechanism of mapping.

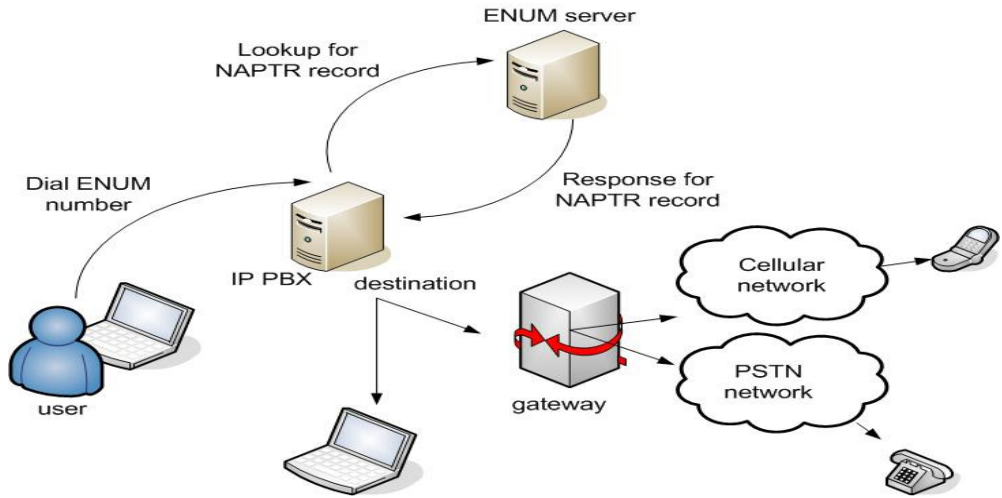

Figure 2. Flow Communication via ENUM

## 2.4. Post Dialling Delay (PDD)

Delay call setup mentioned as Post Dial Delay (PDD). Based on IETF, PDD is a period starting from the last digit dialled up to destination message status (ringing or busy). Based on ITU-T, PDD is an interval between last digit dialled up to the acceptance of ring back tone.

PDD on packet data network especially on SIP VoIP communication can be measured starting from INVITE stage (call request to server) up to destination status tone, such as code 180 (ringing). It can be seen in Figure.3 [2].

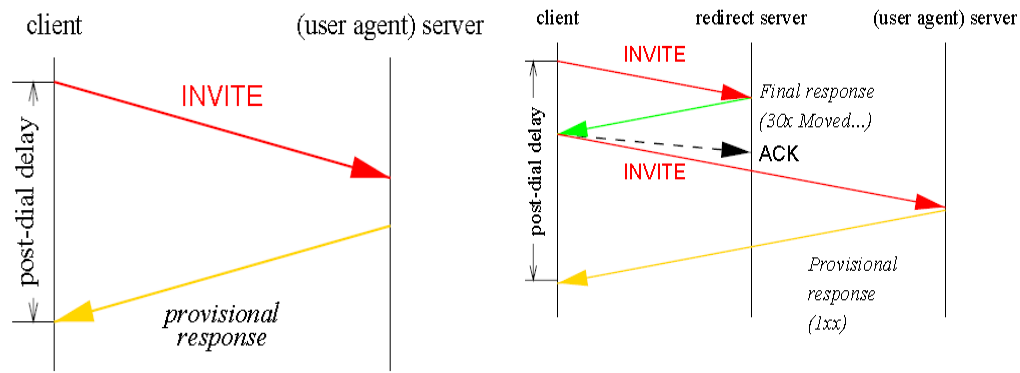

Figure 3. Call setup between SIP client-server

## 3. SYSTEM DESIGN AND IMPLEMTATION

PDD score testing can be performed after the system ensure that it can work properly. System is configured from OpenIMS server, Enum server and Asterisk server for VoIP services.

### 3.1. Network topology of VoIP service

This research is aimed to analyze the relation of SIP terminals especially time server process, PDD time, and CPU utilization. To implement the relation, a scenario of network topology model is applied. This can be seen in Figure 4.

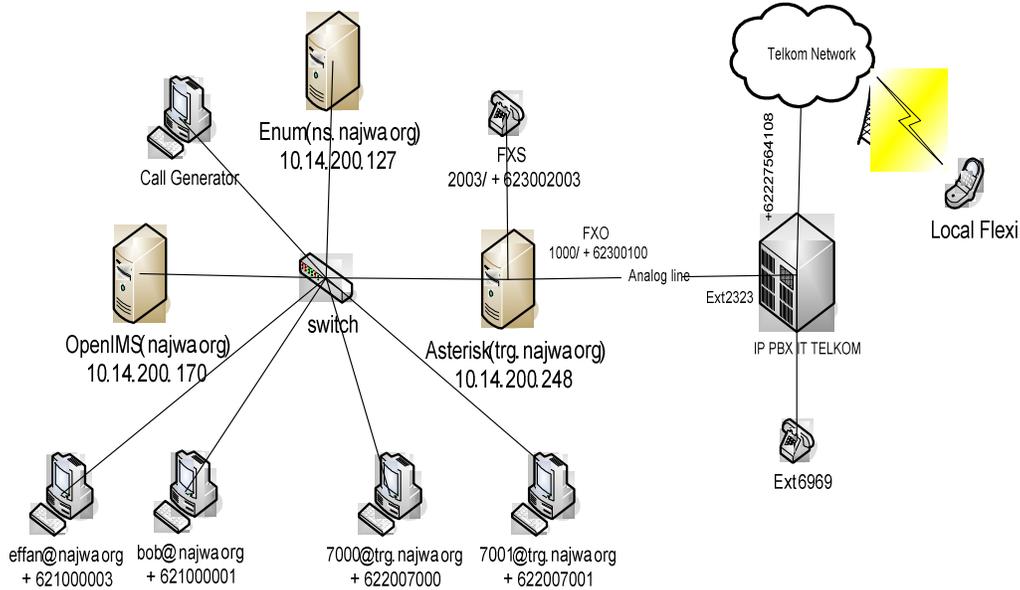

Figure 4. Network Topology

### 3,2, Support of equipment

**Hardware :**
- Processor 1.55 GHz (OpenIMS Server)
- IP PBX Exchange
- Processor 3.0 GHz (Asterisk Sever)
- Processor 1.15 GHz (ENUM Server)
- Computer (Client)
- Switch Ethernet
- Speaker dan Microphone

**Software :**
- Operating System Gentoo 2007 (OpenIMS)
- Operating System Trixbox (Asterisk)
- Operating System Ubuntu 8.04 (ENUM)
- DNS Server (BIND-9.3.2)
- Wireshark Network Protocol Analyzer version-0.99.2

### 3.3. System evaluation

After the system has been installed and well configured, then the next steps are data measurements and analysis.
   a) Time processing in each server measurement.
      For OpenIMS and Asterisk, the measurement will consider time required by SIP signal inside the server before it passed to the destination. While Enum server will be measured

     the required time to serve the query. This measurement performed as an analysis data for PDD measurement [6,7] Measurement's conducted without call traffic/ other queries.
  b) PDD (Post Dial Delay) Measurement
  PDD measurement's performed with scenarios as follows.
  1. Call between OpenIMS without Enum
  2. Call between OpenIMS user via Enum
  3. User call from OpenIMS via Enum to FXS numbers (2003) Asterisk

  Each measurement scenario performed in different call quantity or query per second. Data collections conducted 30 times and averaged.
  c) CPU utilization Measurement
  Measurement's performed for each server for different call quantity or query per second to obtain maximum call value or query per second.

## 4. SYSTEM ANALYSIS

In this chapter we will consider testing and analyzing of implementation results which had been done. The objective of testing and analyzing is to measure OpenIMS and Asterisk Servers performance through Enum server interconnection. As mentioned in previous section, this research will focus on time service analysis in each server, PDD and CPU utilization. Several software have been used such as wireshark, SIPp, Top, and Queryperf-nominum [12,14]
As a reference of PDD maximum value, IETF standard has recommended a recommendation as follows [15]:

Table 1. Standard of PDD maximum value

| Category | PDD without ENUM (second) | WDD with ENUM (second) |
|---|---|---|
| PC to PC | 2.23 | 2.25 |
| PC to PSTN | 3.79 | 4.11 |
| PSTN to PC | 3.41 | 3.95 |
| | | |

### 4.1. Measurement of delay process in Server

The objective of delay process measurement is to measure how much time server can serve a call or query. The value of service time can influence the achieved PDD value. The result of measurement in OpenIMS and Asterisk servers will lead to important information about SIP signal component, which has a long service time..

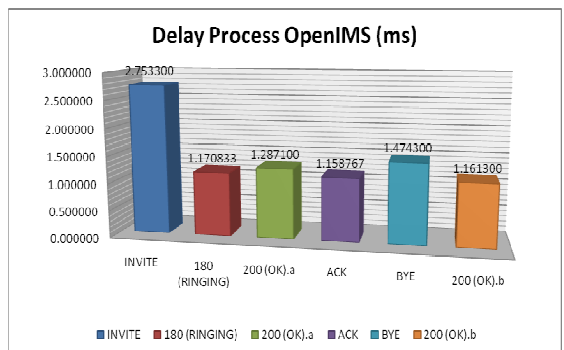

Figure 5. OpenIMS DelayProcess

From figure.5, it can be seen that the invite signal requires longer processing time. When the *server* accept INVITE signal, it will perform some checking. The *Server* will check called user registration status, to fetch the called user IP address and user port and then pass the INVITE Signal to the destination based on called IP address and port user. For the other signal, server does simple checking only; it just acts as proxy that passed the calls. So, the other signals just require short time for server processing.

- Asterisk

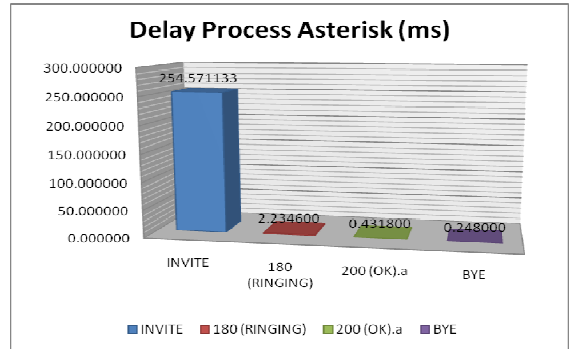

Figure 6. Asterisk Delay Process

Same as the signal process calculation from the OpenIMS, INVITE signal requires longest processing time. This thing happened due to during the INVITE signal acceptance, it will be treated in the same way such as in OpenIMS *server*.

- **Enum**

Based on capture result, the average process in side Enum: **0.345400 ms**, means *server* is capable to proceed a query in a very short time.

- **Comparison**

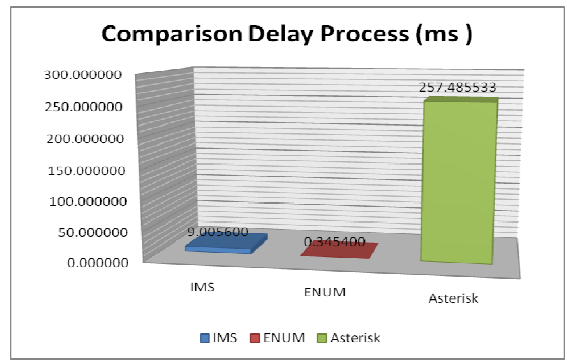

Figure 7. Processing delay comparisons

Based on Figure 7, there is a big difference between Asterisk server process and IMS Server. OpenIMS server requires 9.00 ms for one communication but Asterisk requires 257.48 ms. It means Asterisk server consumes 28.6 times of IMS service time. It can be concluded that the Asterisk operates un-effectively in Call Setup process.

User data in Enum server will be in Mysql database, but in Asterisk it's just sip. conf file, so the time required relatively short during user account checking. Routing process on Open IMS will be recorded on scsf.cfg file (C language used). In the Asterisk server present on extension.cnf

and extension additional. conf (for free PBX). It contains a configuration file. This file will be executed by programming language. To modify dial plan in Asterisk, user may change configuration parameters in this file. It's simple way but will bring the consequences long delay process. Asterisk trixbox has complicated dial plans that will become time consuming for modification.

Enum server has short time process, so it has the capability to handle some queries simultaneously. We will discuss the maximum queries capacity to handle on the next section.

All inserts, figures, diagrams, photographs and tables must be centre-aligned, clear and appropriate for black/white or greyscale reproduction.

### 4.2. PDD measurement and CPU utilization

PDD measurement, start from zero traffic up to certain traffic which cause PDD value higher than IETF standard. Call traffic will be increased 5 calls per second up to PDD value higher than IETF standard. PDD value: time required from the INIVITE signal sent by caller to the acceptance of response ringing signal (180). To collect the PDD values, the wireshark software will be used to capture some data that flows to the caller computer (UAC). Each scenario will be executed for 30 times and averaged.

CPU utilization measurement use Top Software that installed on server. Each traffic capacity value will be observed for the each server maximum CPU Utilization.

- **Result**
- ➢ **First Scenario** (between user OpenIMS without Enum)

INVITE and ringing signal processing time inside the server previously 3.92 ms, measured PDD for 0 traffic is 107.87 ms can be seen at Figure.8. It means that beside server delay, there are some other delay components that influence the PDD time. Client and switch process will add PDD time. Packet assembly and de-assembly in the client side is difficult to observe as well as Switch Reading Process on layer 2.

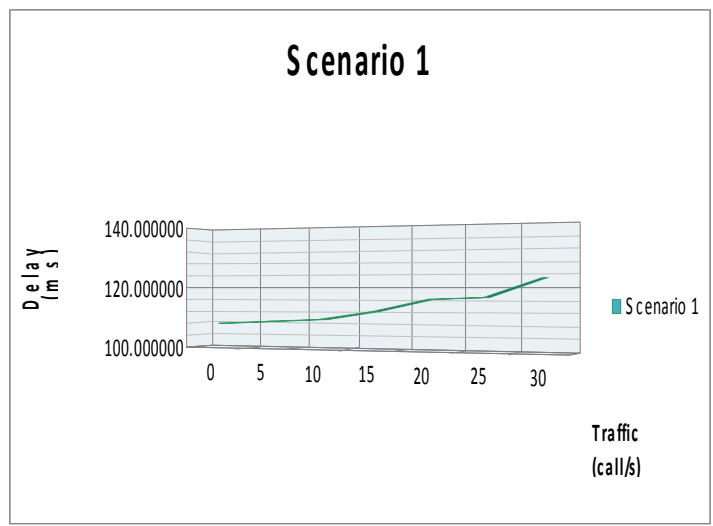

Figure.8   PDD first scenario

The increment of call traffic to server will increase PDD time and will increase processor utility. Up to 30 call/sec PDD's still acceptable (according to IETF standard). Increment of PDD from 0 to 30 calls/sec: less than 20 ms

Table 2. First Scenario PDD measurement

| Traffic (call/sec) | CPU usage (percent) |
|---|---|
| 5 | 10 |
| 10 | 15 |
| 15 | 35 |
| 20 | 50 |
| 25 | 65 |
| 30 | 78 |
| 35 | 100 (overload) |

For a traffic call more than 30 call/sec, OpenIMS server can not handle the call. Monitoring result on SIP's showed so many Re-trans are occurred, Timeout and Unexpected-Msg. It shows the server status: overload and unable to serve as be shown in Table.2. PDD measurement, call setup will be too long. This monitoring facts showed by wireshark client do re-trans over and over again. So for the call traffic above 30 call/sec, server's unable to serve calls. If this situation forced to handle more than 30 call/sec, server will be downed (restart).

**Second Scenario** (between OpenIMS user through Enum).

From the measurement result, increment of Enum components during call setup will increase PDD value. For traffic up to 30 call/s, the PDD value still comply with IETF standard, that can be seen in Table.3

Table 3. Second scenario PDD measurement

| OpenIMS-ENUM-OpenIMS | |
|---|---|
| Traffic (call/sec) | PDD (msec) |
| 0 | 108.8 |
| 5 | 109.5 |
| 10 | 110.4 |
| 15 | 114.1 |
| 20 | 116.4 |
| 25 | 118.6 |
| 30 | 133.1 |

This is caused by faster process in Enum server; PDD value with Enum only increase 0-10 ms, if compared without Enum. It's can be seen in Figure 9 below.

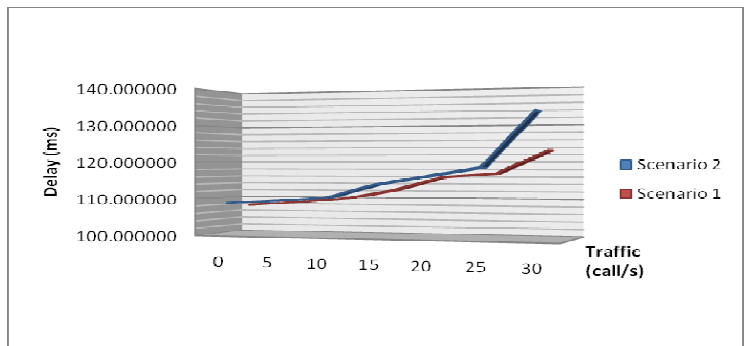

Figure 9. Comparison of PDD first and second scenario

Table 4. CPU Utilization measurement through Enum

| OpenIMS-ENUM-OpenIMS | |
|---|---|
| Traffic (query/sec) | CPU usage (percent) |
| 5 | 0.3 |
| 10 | 0.3 |
| 15 | 0.3 |
| 20 | 0.3 |
| 25 | 0.7 |
| 30 | 0.7 |
| 35 | 0.7 |

From the above results, OpenIMS server only has the capability for simultaneous traffic call 30 call/sec, so the data traffic for this scenario is only limited up to 30 call/sec. On call traffic 30 call/sec, the CPU utilization in Enum server only 0.7% as in Table.4; this figure is more than maximum value. The maximum query per second can be processed using software called queryperf. From the measurement result with queryperf, it can be shown, that the Enum server can reached 8156.87 query/second.

➢ **Third Scenario** (from the user OpenIMS to FXS Asterisk through Enum)

The third scenario can show results of PDD which is still fulfill IETF standard. In this scenario the PDD value uses is higher than other scenarios. This condition is caused by time process in Asterisk server for an invite signal consume more time is 254.5 ms. The comparison of all scenarios for PDD value can be seen in Figure 10 bellow

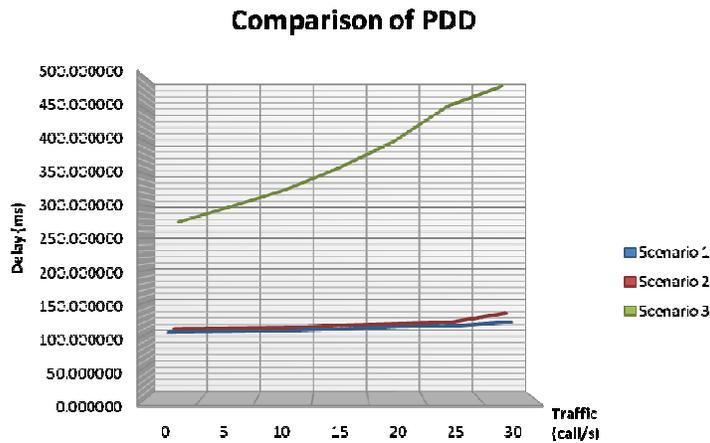

Figure 10. Comparison of PDD for all Scenarios

Table.5 shows that for call traffic 35 call per second, CPU utilization in Asterisk can not reach maximum value. The next test is to find maximum call with using SIP software. When the arrival call higher than 55 call per second, CPU utilization will be on an overload condition. From this test we found that Asterisk server can only handled maximum 55 call per second.

Table 5. CPU Utilization Asterisk

| OpenIMS-ENUM-Asterisk ||
| Traffic (call/sec) | CPU usage (percent) |
| --- | --- |
| 35 | 51 |
| 40 | 60 |
| 45 | 69 |
| 50 | 79 |
| 55 | 92 |
| 60 | 100 (overload) |

Figure.11 shows comparison of CPU utilization for all servers with attention to load traffic for each server.

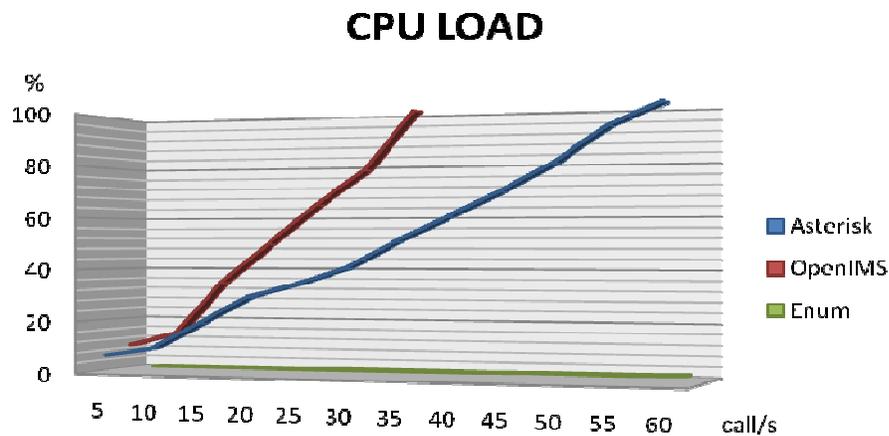

Figure 11. Comparison of CPU Utilization

On this implementation, OpenIMS server only capable maximum service 30 call per second, but for Asterisk server can reaches 60 call per second. The Asterisk server can handle a large number of call, because it uses high rate of processor – 3.0 GHz; OpenIMS uses only processor rate-1.55 GHz. Another server for Enum with rate of processor-1.15 GHz has the capability to handle untill 8156 query per second.

## 5. CONCLUSIONS

Based on the implementation, tests and analysis , we can conclude some results as follows :
1. Interconnection between OpenIMS and Asterisk through the Enum server has been done with good results. This interconnection shows that it is possible that two users can be connected by using Asterisk as a Gateway in a circuit network.
2. For call set up phase, the processing time in Asterisk server is longer than in openIMS server. The Asterisk server focuses more for user in easy to handle fixed routing process, while OpenIMS focuses on the rate of routing process.
3. Some cases affecting PDD values are server call time processing and the amount of traffic flow to the server. From all scenarios, the result of PDD value that lies under the range of IETF standard is 2.3 second. The average PDD maximum value is 493.66 second, found in scenario-3, with the amount of background traffic 30 call per second. In this scenario, a call will pass all three servers : Open IMS, Enum and Asterisk.

4. The amount of arrival traffic or query which is capable to be serviced by the server, depending on the server types and the specification of the computer used in the system.

## ACKNOWLEDGEMENTS

The authors would like to thanks the ITTelkom and YPT in Bandung for their financial support. They also express their sincere thanks and appreciation to all colleagues at ITTelkom for their kind help and usefull suggestions.

Short Biography

**Rendy Munadi** received his Dr from Indonesia University . He is a senior lecturer of ITTelkom Bandung-INDONESIA and he is presently as Director of postgraduates program. Dr Rendy Munadi has served on the program committee of several conference. He is current research in the area of Next Generation Network, IMS, wireless network, IP/MPLS network, routing management and protocol SIP, H323 etc. , e-mail: rnd@ ittelkom.ac.id).

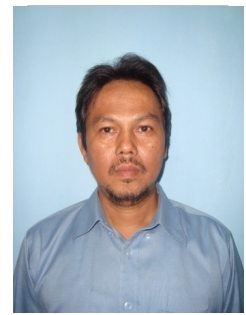

**Asep Mulyana** received his MT from Bandung Institute of Technology-Indonesia. He has experienced in the field on switching system (exchange), experienced as trainer on various technology of switching system, signalling system and telecommunication networks in the Training Centre PT. TELKOM Bandung. Currently he is a lecturer on Switching Technique, Telecommunication Networks, Traffic Engineering, Access Networks, Signalling System and Next Generation Networks (NGN). At this moment he is interest and concern to research on NGN and IMS

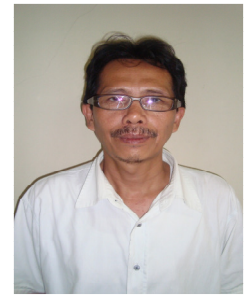

**R. Rumani Mangkudjaja**; Department of Electrical and Communication, Telkom Institute of Technology, Bandung 40257, INDONESIA. R. Rumani Mangkudjaja was born in Balikpapan, INDONESIA, on 1947, April 4. He received the BS degrees in Education from Institute of Teaching and Education, Bandung, Indonesia, in 1978, and Electrical Engineering from Maranatha Christian University, Bandung, Indonesia in 1981, and the MS degree in Communications from the George Washington University, Washington, DC, USA, in 1994.
Currently, he is a lecturer of Electrical and Communication Engineering at Telkom Institute of Technology, Bandung, Indonesia, since 1994.
His research interests include computer and communication networks, next generation networks, ISDN, intelligent networks, queuing theory, traffic theory, and related fields.
.

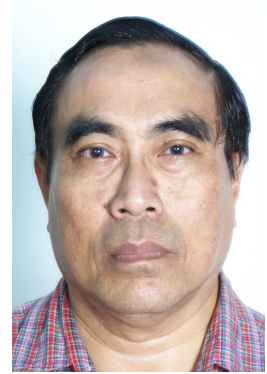